\documentclass{ws-ijmpd}
\usepackage[super,compress]{cite}
\usepackage{graphicx}
\begin{document}


%
\catchline{}{}{}{}{}
%

\title{ASTRODYNAMICAL SPACE TEST OF RELATIVITY USING OPTICAL DEVICES I (ASTROD I) - MISSION OVERVIEW
}

\author{HANNS SELIG, CLAUS L\"{A}MMERZAHL
}

\address{ZARM, University of Bremen, Am Fallturm\\
28359 Bremen,
Germany
\\
hanns.selig@zarm.uni-bremen.de, claus.laemmerzahl@zarm.uni-bremen.de}

\author{WEI-TOU NI}

\address{CGC, Department of Physics, National Tsing Hua University, Hsinchu, Taiwan, 300, ROC\\
weitou@gmail.com}

\maketitle

\begin{history}
\received{Day Month Year}
\revised{Day Month Year}
\end{history}

\begin{abstract}
ASTROD I is the first planned space mission in a series of ASTROD missions for testing relativity in space using optical devices.
The main aims are: (i) to test General Relativity with an improvement of three orders of magnitude compared to current results,
(ii) to measure solar and solar system parameters with improved accuracy, (iii) to test the constancy of the gravitational constant
and in general to get a deeper understanding of gravity. The first ideas for the ASTROD missions go back to the last century
when new technologies in the area of laser physics and time measurement began to appear on the horizon. ASTROD is a mission concept that is 
supported by a broad international community covering the areas of space technology, fundamental physics, high performance laser and clock technology and drag free control.
While ASTROD I is a single-spacecraft concept that performes measurements with pulsed laser ranging between the spacecraft and earthbound laser ranging stations,
ASTROD-GW is planned to be a three spacecraft mission with inter-spacecraft laser ranging. ASTROD-GW would be able to detect gravitational waves at frequencies below the  eLISA/NGO bandwidth. As a third step Super-ASTROD with larger orbits could even probe primordial gravitational waves.
This article gives an overview on the basic principles especially for ASTROD I.

\end{abstract}


\ccode{PACS numbers:}


\section{Introduction}	

The ASTROD (Astrodynamical Space Test of Relativity using Optical Devices) mission series aims at high precision measurements in the interplanetary space for the determination of several quantities in the areas of General Relativity and Fundamental Physics as well as solar and solar system research \cite{A01}. The main goals are improvements in the determination of the relativistic parameterized-post-Newtonian (PPN) parameters $\beta$ and $\gamma$ (Eddington parameter), the solar and solar system parameters and the variation of the gravitational constant. Gravitational wave detection is a further goal for the multi-spacecraft missions ASTROD-GW \cite{ASTROD-GW,A02} and Super-ASTROD \cite{A03}. In the frame of this overview article on the single-spacecraft ASTROD I mission concept the specific mission goals, the basic mission concept as well as several technical aspects of the mission are presented. More detailed descriptions of the different mission aspects are given in Refs.~\refcite{ASTROD,ASTROD-alt}.

\begin{figure}[pb]
\centerline{\psfig{file=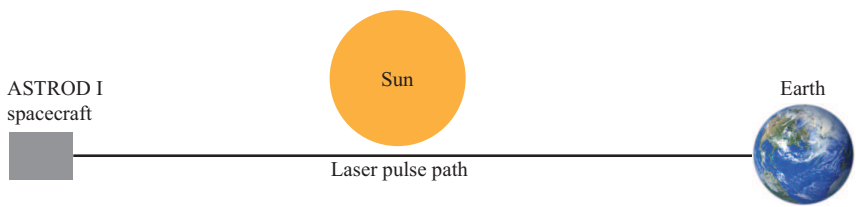,width=12.0cm}}
\vspace*{8pt}
\caption{Constellation scheme for the measurement of the Shapiro time delay for the determination of the PPN-parameter $\gamma$ (Eddington parameter). \label{f1}}
\end{figure}

ASTROD I is a single spacecraft operated in drag-free-mode with a laser link to the Earth (ground based laser stations) using pulsed laser ranging with state of the art measurement techniques. The main payload components are a two color pulsed laser system, a laser pulse receiving unit, an event timer with a corresponding atomic clock for precise timing measurements with a timing accuracy of 3 ps and a high performance drag free control system containig an inertial sensor and several micro-Newton thrusters for a precise attitude and orbit control. To achieve the mission goals the spacecraft needs to follow a pure gravitational orbit (geodesic) which means that all disturbing forces have to be compensated by the drag free system to a high precision. The ideal orbit of ASTROD I can be reached with the help of Venus encounters using the swing-by-technique. Especially the period when the spacecraft is located on the opposite of the Sun (viewed from the Earth - Fig.~\ref{f1}) is very important for the measurement of the Shapiro time delay which can be used for the determination of the Eddington parameter $\gamma$. Due to the special orbit with Venus encounters this leads to a minimum Sun distance of less than 0.5 Astronomical Units (AU) which is a challenge for the mission design in terms of thermal load for the spacecraft and in terms of optical setup and sunlight shielding/filtering. The most recent simulations for possible orbits with corresponding Venus encounters show that there are several launch windows for ASTROD I. Detailed calculations have been carried through with a launch date in 2021 and a mission duration of 1200 days\cite{ASTROD}. In principle there are many possible launch windows due to the fact that the orbital periods of Earth and Venus are short - especially compared to the orbital periods of the planets in the outer solar system. Therefore there is not a strong constraint on the year of launch.

In 2010 ASTROD I was proposed for the ESA Cosmic Vision 2015-2025 M-Class program by an international researcher team. ASTROD I was selected as one of the final 14 candidates out of nearly 50 proposals which shows the genaral interest in the mission concept. Finally ASTROD I was not selected yet.

\section{Mission goals overview}

\begin{table}[ph]
\tbl{Mission goals for the ASTROD I mission (for 1200 days).}
{\begin{tabular}{@{}cccc@{}} \toprule
Effect/Quantity & Present accuracy & projected accuracy  \\ \colrule
PPN parameter $\beta$ & 2 $\times$ $10^{-4}$ & 6 $\times$ $10^{-6}$  \\
PPN parameter $\gamma$ (Eddington parameter) & 4.4 $\times$ $10^{-5}$ & 3 $\times$ $10^{-8}$  \\
(dG/dt)/G & $10^{-12}$ $\text{yr}^{-1}$ & $10^{-14}$ $\text{yr}^{-1}$ \\
Anomalous Pioneer acceleration $A_a$  & (8.74 $\pm$ 1.33) $\times$ $10^{-10}$ m/$\text{s}^2$ & 6 $\times$ $10^{-16}$ m/$\text{s}^2$ \\
Determination of solar quadrupole moment parameter $J_2$ & (1.82 $\pm$ 0.47) $\times$ $10^{-7}$ & 1 $\times$ $10^{-9}$ \\
Solar angular momentum via solar Lense Thirring Effect & 0.1 (Earth Lense-Thirring Effect) & 0.1 \\
Planetary masses and orbit parameters & (depends on object) & 1-3 orders better \\
Asteroid masses and densities & (depends on object) & 2-3 orders better\\ \botrule
\end{tabular} \label{ta1}}
\end{table}

There are three main scientific goals. The first goal is to test relativistic gravity and the fundamental laws of spacetime with an improvement of three orders of magnitude in sensitivity, specifically, to measure the PPN parameters $\gamma$ (Eddington light-deflection parameter; for general relativity, it is 1) via measurements of the Shapiro time delay to 3 $\times$ $10^{-8}$, $\beta$ (relativistic nonlinear-gravity parameter; for general relativity, it is 1) to 6 $\times$ $10^{-6}$ and others with significant improvement; and to measure the fractional time rate of change of the gravitational constant (dG/dt)/G with two orders of magnitude improvement. The Pioneer Anomaly could meanwhile be explained as an anisotropic thermal radiation effect \cite{pioneer}. Therefor a further investigation on the Pioneer Anomaly is not of strong interest anymore. 
However, relativistic MOND (MOdified Newtonian Dynamics) theories \cite{A04} are of current interests as they are alternatives to dark matter theories. MOND theories have effects similar to the Pioneer effect in the solar system. Testing MOND using LISA Pathfinder \cite{A05} has been considered and analyzed \cite{A06,A07}. With ASTROD I precision, MOND effects on ASTROD I need to be studied, and possible tests on MOND theories would then be evaluated\footnote[1]{The MOND effects are about $10^{-10}$ m$\text{s}^{-2}$. Although, ASTROD I will not be made to pass through the Sun-Earth Saddle point like LISA Pathfinder, it is sensitive to long term acceleration deviations much below this level. If studies show that the MOND deviations from general relativity for gravity accelerations above the $10^{-10}$ m$\text{s}^{-2}$ threshold are significant, ASTROD I would complement LISA Pathfinder to test various MOND theories in this sense.}.

The second goal is to initiate a revolution of astrodynamics with laser ranging in the solar system. With mm precision, the astrodynamics and ephemeris of the solar-system bodies would be improved by orders of magnitude.

The third goal is to increase the sensitivity of the determination of solar, planetary and asteroid parameters by 1 - 3 orders of magnitude. In this context, the measurement of the solar quadrupole moment parameter $J_2$ will be improved by two orders of magnitude, i.e., to $10^{-9}$. 
 The mission goals for ASTROD I are listed in Table~\ref{ta1}). These goals are based on  orbit simulation and assumptions for a timing uncertainty of 3 pico-seconds (ps) and a drag-free uncertainty of 3 $\times$ $10^{-14}$ m\,s$^{-2}$ $\text{Hz}^{-1/2}$ at 0.1 mHz. The timing uncertainty has been achieved in T2L2 on board Jason 2 \cite{ASTRODref6,ASTRODref7,A08}. The drag-free requirement is comparable to the actual goal of LISA Pathfinder \cite{A05}. See section 5 for more discussions.
 
 In the area of solar and galactic physics ASTROD I can also be used for measuring solar energetic particles (SEPs) and galactic cosmic rays (GCRs), with corresponding applications to space weather \cite{ASTROD,A09,A10}. The radiation detectors onboard the spacecraft for monitoring test-mass charging of the inertial drag-free sensor can be used for this purpose.  
 
 Like LLR, the scientific data are ranges between ASTROD I and laser stations on earth as functions of time. All goals can (must) be extracted from these ranges with fittings. No payload system could be omitted without affecting the range measurement except the possibility of omitting the radiation monitor which is only a small part of the ASTROD I payload related to the inertial sensor and which is good for space weather applications.

\section{Spacecraft orbit selection}
\begin{figure}[bh!]
\centerline{\psfig{file=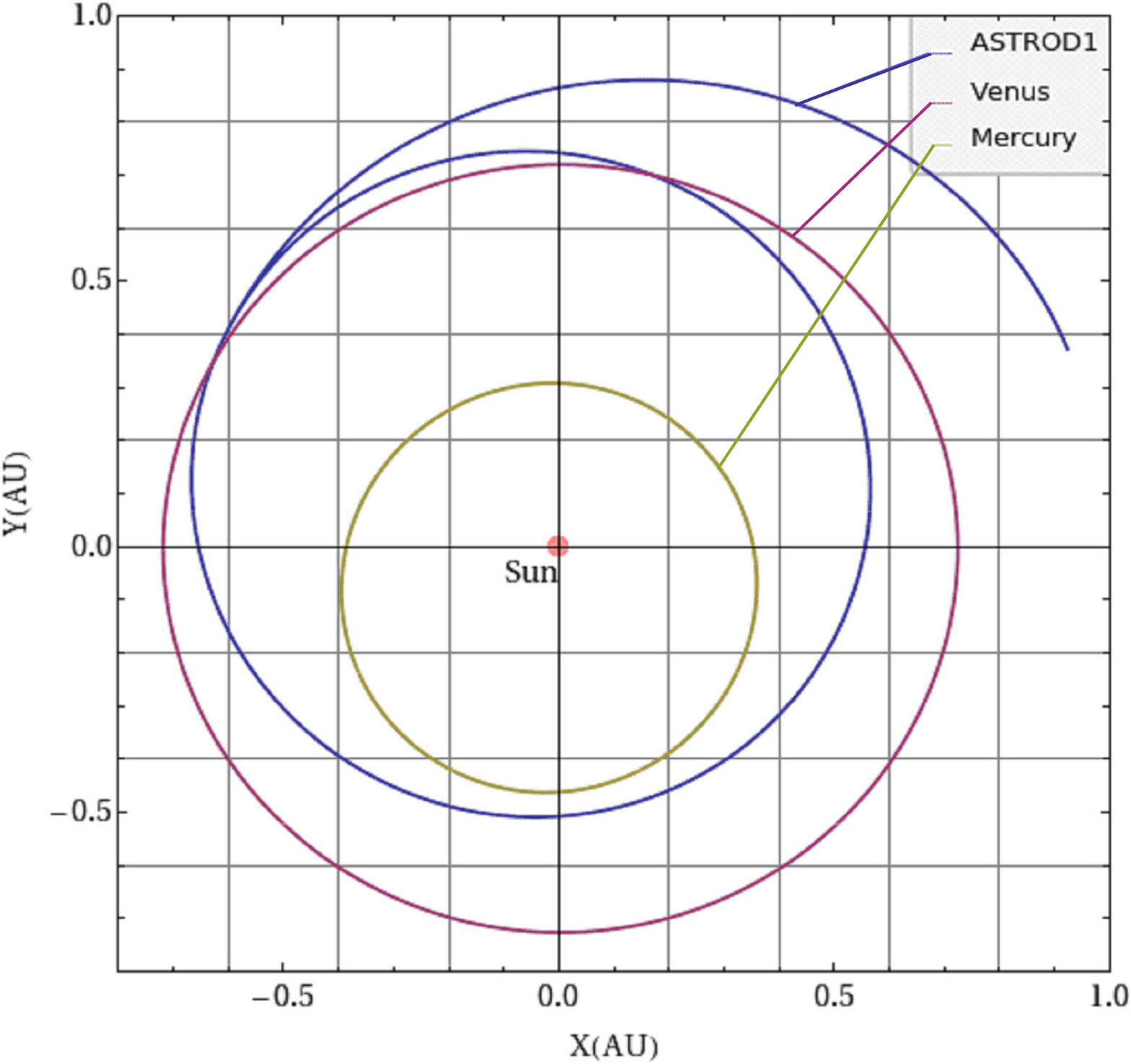,width=7.0cm}}
\vspace*{8pt}
\caption{A typical orbit in heliocentric ecliptic coordinate system. \label{f2}}
\end{figure}

\begin{figure}[h]
\centerline{\psfig{file=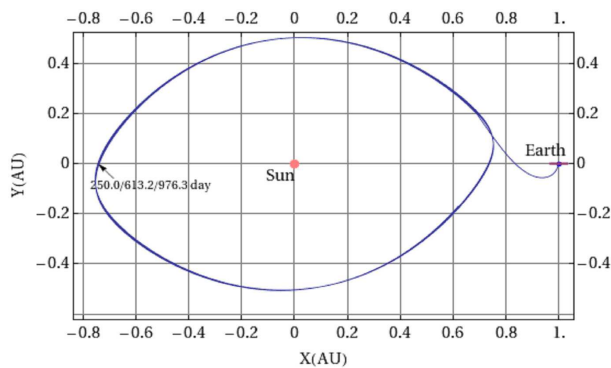,width=8.0cm}}
\vspace*{8pt}
\caption{The 2021 orbit in the Sun-Earth fixed frame. \label{f3}}
\end{figure}
Besides the technical details of ASTROD I (spacecraft and payload) the success of the mission depends on the selection of an appropriate orbit to reach the mission goals.
Some effort has been made to simulate possible orbits including planetary swing-bys in order to find adequate solutions \cite{ASTROD}. The swing-bys are intended to be geodesic and can be used for mapping Venus gravity. The fuel requirement is minimal with only drag-free thrusters needed. 
One simulation with a launch in 2021 and a Venus encounter 112 days after launch and three oppositions (spacecraft, Sun and Earth in one line of sight) - after 250, 613 and 976 days after launch - gives a good example for such an orbit \cite{9}. For the choice of the best orbit a trade-off is needed between minimising the time to reach the far side of the Sun because of the risk to the mission from aging of equipment.

The simulated launch-2021 orbit for the ASTROD I spacecraft is shown in Fig.~\ref{f2}, the corresponding orbit in the Sun-Earth fixed frame in Fig.~\ref{f3}. The Venus swing-by strategy is also useful for ASTROD-GW mission design and optimization \cite{A11}.

\section{The ASTROD I spacecraft}
\begin{figure}[h]
\centerline{\psfig{file=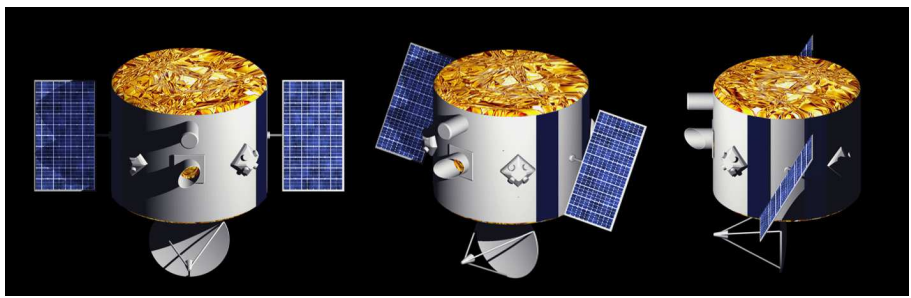,width=12.0cm}}
\vspace*{8pt}
\caption{Preliminary design of the ASTROD I spacecraft (design by H. Selig, ZARM). \label{f4}}
\end{figure}
ASTROD I is planned as a cylindrical spacecraft with a diameter of 2.5 meters and a hight of 2 meters.
The total mass including a 100 kg apogee motor with liquid fuel is 490 kg. The spacecraft is stabilized around 3 axes by a drag free system.
The total electrical power (including payload) is estimated to be around 390 W.
In order to protect the payload from thermal load of sunlight - especially at close distances to the Sun - the spacecraft will have to be equipped with passive thermal insulation systems - e.g.: multi-layer-insulation (MLI), and perhaps even with active thermal control devices like thermel Louvers. The minimum requirement for the thermal stability of the payload is a temperature variation of less than $\pm$ 1 K. Due to the variation of the distance between the spacecraft and the Sun (0.4 to 0.8 AU) the thermal influx varies by a factor of 4 \cite{ASTROD}. This will have to be taken into account for the detailed spacecraft design in terms of thermal control.
The needed electrical power of 390 W will be generated by solar cells. One possibility is to cover the spacecraft surface with the needed area of solar cells. This could create a thermal problem due to the limited efficency of solar cells. Assuming an efficency of 0.2, most of the lost power will be transformed into heat. In order to prevent the spacecraft from heating up, the solar cells could also be placed on solar panels (see Fig.~\ref{f4}).
For the radio communication with the Earth the spacecraft is equipped with a 1.3 m diameter X-band antenna and low gain S-band antennas. 
During the whole mission the ASTROD I spacecraft will have to be pointed towards the earth to keep the laser link running. Therefore the high gain antenna does not need to be tracked as long as the high gain antenna is pointed parallel to the laser optics which simplifies the setup.

\section{The ASTROD I payload}

The ASTROD I payload consists of the following components:

\begin{itemlist}
\item 300 mm Cassegrain telescope
\item Optical bench
\item Two color pulsed Nd:YAG-laser system (532 nm, 1064 nm)
\item Single photon detecting photodiodes
\item Event timer
\item Cesium/Rubidium clock
\item Drag-free inertial sensor
\item Drag-free micro-Newton-thrusters
\end{itemlist}

\begin{figure}[h]
\centerline{\psfig{file=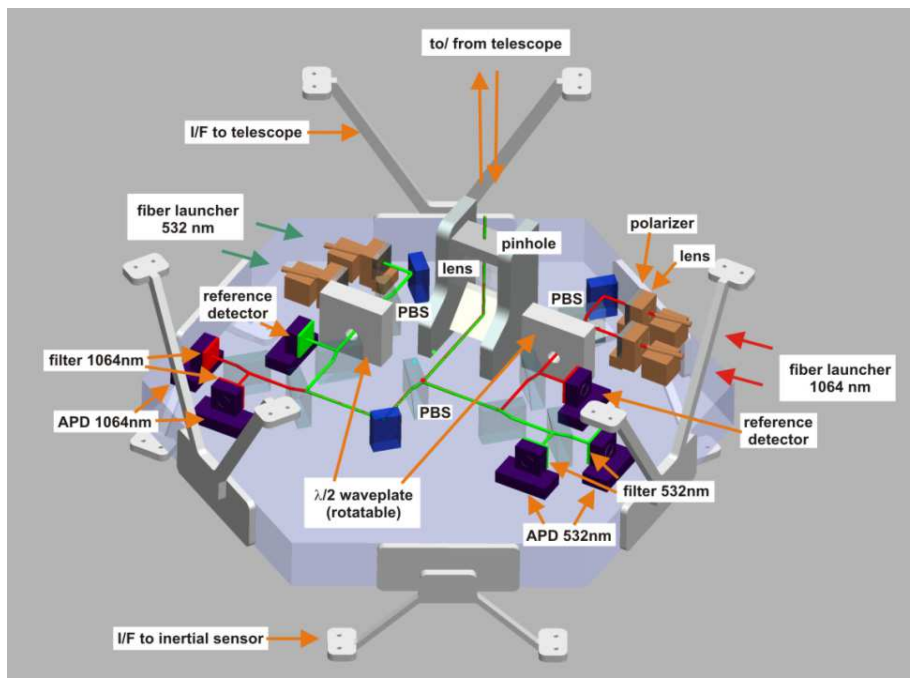,width=12.0cm}}
\vspace*{8pt}
\caption{ASTROD I optical bench \cite{ASTROD}. \label{f5}}
\end{figure}

The Cassegrain telescope is used for transmitting and receiving the laser pulses from/to Earth/spacecraft.
The optical bench consits of several optical components for processing the laser light inside the spacecraft. The main purpose of the optical bench is to seperate the incoming/outgoing laser pulses by polarisation- and color-filtering. Fig.~\ref{f5} shows the preliminary design of the optical bench. Detailed descriptions are given in Refs.~\refcite{ASTROD,ASTROD-alt}.
The laser system onboard the spacecraft generates laser pulses with two wavelengths (532 and 1064 nm). The reason for this two-color setup is the need for a correction of the effect of the air column of the atmosphere. With two lasers with sufficiently different wavelengths it is possible to measure and substract the effect \cite{ASTROD-alt}.
It is sufficient as demonstrated already by SLR and LLR. The pulse width is 50 ps, the repetition rate 100 Hz and the pulse energy around 10 mJ. 

The specific type of the Nd:YAG laser system is not yet fixed. The choice depends on the mass and performance and on the availability of space qualified units.

Single photon detectors are intended to be used for the laser pulse reception onboard the spacecraft. For the required 3 ps timing accuracy a precise event timer is needed.
The 3 ps timing accuracy is already achieved by the T2L2 (Time Transfer by Laser Link) event timer onboard Jason 2 \cite{ASTRODref6,ASTRODref7,A08}.

The emitting times and receiving times
will be recorded by a cesium atomic clock. For a ranging uncertainty of 0.9 mm
in a distance of 2.55 $\times$ $10^{11}$ m (1.7 AU), the laser/clock frequency needs to be
known to one part in $10^{14}$ by comparison with ground clocks over a period of
time. Stability to 6 $\times$ $10^{-14}$ in 1,000 s (round-trip time) is required.
Interplanetary pulse laser ranging(both up and down) was demonstrated by MESSENGER, using its laser
altimeter in 2005 \cite{ASTRODref18}. The technologies needed for a dedicated mission using
interplanetary pulse laser ranging with millimetre accuracy are already mature.

A drag-free sensor with a free falling test mass is necessary as a reference for the pure gravitational motion of the spacecraft and micro-Newton thrusters are needed to compensate disturbing forces acting on the spacecraft in a closed loop control system. The drag-free
requirement for ASTROD I is relaxed by one order of magnitude as compared
with that of LISA (Laser Interferometer Space Antenna for gravitational wave
detection) \cite{ASTRODref13}. The drag-free technologies under development for LISA
Pathfinder \cite{ASTRODref14,A05} will largely meet the requirement of ASTROD I. Initially the FEEP technology (Field-Emission Electric Propulsion) seemed to be the best choice for micro-Newton thrusters for missions like LISA Pathfinder, MICROSCOPE and the ASTROD missions. Due to technical problems during the development of the FEEP technology the cold gas technology has also been taken into acount for these missions. The GAIA mission will carry cold gas thrusters for the AOCS (Attitude and orbit control system)\cite{GAIA}. MICROSCOPE and LISA Pathfinder will be equipped with cold gas thrusters based on the GAIA thrusters. The main disadvantage of cold gas thrusters compared to FEEPs is the higher mass per $\Delta$v. The total mission duration is limited by the amount of propellant stored in the tanks. Therefore the FEEP technology would be preferred if it is available soon enough for ASTROD I.

The successful functioning of the accelerometer on board GOCE \cite{ASTRODref15} reassured the
development for achieving improved inertial sensors. ASTROD-GW, DECIGO Pathfinder and DECIGO also require drag-free technologies for their performances \cite{A02,A12}.

Sunlight shielding is a common technology which needs to be adopted to
the special requirements for ASTROD I (see next section). The overall technology readiness level of the ASTROD I payload components will be improved during the further development.

\section{Signal detection and sunlight filtering}

For the measurement of the post-Newtonian parameter γ (Eddington light deflection
parameter) near the opposition, i.e., 1.7 AU = 255 million km away
from Earth, the spacecraft telescope will pick up only about $10^{-13}$ of the
power emitted. For the wavelength of 1064 nm that would mean a pulse of
$10^{-15}$ J or 5,000 photons \cite{ASTROD-alt}. Since Avalanche Photodiodes (APD) are used for single photon
detection, a beam attenuation may be needed before it enters the detector.
At the same time, the sunlight will, at opposition, shine with 400 W into
the spacecraft telescope and extreme care must be taken
to reduce that sunlight to a level that the laser signal is dominant. 
Three different measures are planned for the reduction. (i) Spatial filtering by placing a pinhole plate in the focal plane of the telescope, (ii) spectral filtering with narrow bandwidth dielectric filters and (iii) temporal filtering (10 ns window). 
The remaining solar photons will be sufficiently less
than the laser photons at the photo detector and the laser pulse signals can be
detected. Especially the spectral narrow band filtering for two wavelengths (in one filter) needs some more investigation. When there is a Phase A study, laboratory implementation of the sunlight filtering system needs to be demonstrated.

\section{Study on a Combined ASTROD I and OPTIS mission}

OPTIS is a satellite-based test of Special and General Relativity \cite{A13}. The tests are based on ultrastable optical cavities, lasers, an atomic clock and a frequency comb generator. OPTIS projects a Michelson-Morley test, a Kennedy-Thorndike test and a test of the universality of the gravitational redshift by comparison of an atomic clock with an optical clock. For OPTIS, a laser link to the ground for comparison with ground clocks is much desired. In the early version of ASTROD I, interferometric ranging is included. For the interferometric ranging, the frequency of the laser offset-phase-locked to the incoming light can be measured by comparison with a harmonic frequency generated by an optic comb using a standard input frequency from the Cs clock on the spacecraft. It would be natural to combine these two versions of ASTROD I and OPTIS (as to acronym, it could be ASTROPTIS, ASOP,…) and go to deep space. Michelson-Morley experiments could gain much sensitivity from a deep space mission. Redshift comparison would be good also. Kennedy-Thorndike tests may lose some sensitivity. However, deep space interferometric ranging will be tested. More study regarding to costs is needed.

\section{Conclusion}

The field of the experimental gravitational physics stands to be revolutionized by the advancements in several critical technologies, over the next few years. These technologies include deep space drag-free navigation and interplanetary laser ranging. A combination of these serves as a technology base for ASTROD I. ASTROD I is a solar-system gravity survey mission to test relativistic gravity with an improvement in sensitivity of three orders of magnitude, improving our understanding of gravity and aiding the development of a new quantum gravity theory; to measure key solar system parameters with increased accuracy, advancing solar physics and our knowledge of the solar system; and to measure the time rate of change of the gravitational constant with an order of magnitude improvement and probing dark matter and dark energy gravitationally. This will be the beginning of a series of precise space experiments on the dynamics of gravity. The techniques are becoming mature for such experiments.

\end{document}